\documentclass{SCGE}

\begin{document}

\begin{picture}(0,0){\rm
\put(0,-20){\makebox[160truemm][l]{\bf {\sanhao\raisebox{2pt}{.}}
Article  {\sanhao\raisebox{1.5pt}{.}}}}}
\put(0,-34){\jiuwuhao {\textcolor[rgb]{0.5,0.5,0.5}{\sf 
}}}
\end{picture}

\def\bm{\boldsymbol}

\def\dl{\displaystyle}
\def\du{\end{document}}
\def\d{{\rm d}}
\def\e{{\rm e}}
\def\i{{\rm i}}
\def\pi{{\uppi}}

\Year{2017} %
\Month{??} %
\Vol{??} 
\No{?} 
\BeginPage{1} 
\AuthorMark{{\rm Ze Zhao}, et al.}  
\AuthorMarkCite{{\rm Ze Zhao, \& Shuang Wang} } 
\DOI{??} 
\ArtNo{??}

\title[Diagnosing HDE models]
{Diagnosing holographic type dark energy models with the Statefinder hierarchy, composite null diagnostic and $w-w'$ pair}

\author[1,2]{Ze Zhao}{}
\author[3]{Shuang Wang}{wangshuang@mail.sysu.edu.cn}

\address[{\rm1}]{CAS Key Laboratory of Theoretical Physics, Institute of Theoretical Physics, Chinese Academy of Sciences, Beijing 100190, P. R. China;}
\address[{\rm2}]{School of Physical Science, University of Chinese Academy of Science, No.19A Yuquan Road, Beijing 100049, P. R. China;}
\address[{\rm3}]{School of Astronomy and Space Science, Sun Yat-Sen University, Guangzhou 510275, P. R. China}

\maketitle \vspace{-3.5mm}{\footnotesize\begin{center} 
\end{center}}\vspace*{-5mm}

\begin{center}
\rule{16.5cm}{0.4pt}
\parbox{16.5cm}
{\begin{abstract}
The main purpose of this work is to distinguish various holographic type dark energy (DE) models, including the $\Lambda$HDE, HDE, NADE and RDE model, by using various diagnostic tools. The first diagnostic tool is the Statefinder hierarchy,
in which the evolution of Statefinder hierarchy parmeter $S^{(1)}_3(z)$ and $S^{(1)}_4(z)$ are studied.
The second is composite null diagnostic (CND), in which the trajectories of $\{S^{(1)}_3, \epsilon\}$ and $\{S^{(1)}_4, \epsilon\}$ are investigated, where $\epsilon$ is the fractional growth parameter. The last is $w-w'$ analysis, where $w$ is the equation of state for DE and the prime denotes derivative with respect to $ln a$. In the analysis we consider two cases: varying current fractional DE density $\Omega_{de0}$ and varying DE model parameter $C$. We find that:
(1) Both the Statefinder hierarchy and the CND have qualitative impact on $\Lambda$HDE, but only have quantitative impact on HDE.
(2) $S_4^{(1)}$ can lead to larger differences than $S_3^{(1)}$, while the CND pair has a stronger ability to distinguish different models than the Statefinder hierarchy.
(3) For the case of varying $C$, the $\{w, w'\}$ pair has qualitative impact on $\Lambda$HDE; for the case of varying $\Omega_{de0}$, the $\{w, w'\}$ pair only has quantitative impact; these results are different from the cases of HDE, RDE and NADE, in which the $\{w, w'\}$ pair only has quantitative impact on these models.
In conclusion, compared with HDE, RDE and NADE, the $\Lambda$HDE model can be easily distinguished by using these diagnostic tools.
\end{abstract}}
\end{center}\vspace*{-0.6cm}

\begin{center}
\parbox{16.5cm}
{\bf\jiuhao Key words: dark Energy, cosmology, cosmological constant}
\end{center}

\begin{center}
{\PACS{\rm 95.36.+x,  98.80.¨Ck, 98.80.Es.}}
\CITA    
\end{center}

\textwidth=178truemm \textheight=236truemm

\wuhao\vspace*{1.5mm}

\begin{multicols}{2}

\renewcommand{\baselinestretch}{1.08} \baselineskip 12.2pt\parindent=10.8pt

\renewcommand{\thefootnote}


\section{Introduction}

Various astronomical observations, such as Type Ia Supernovae (SN~Ia) \cite{Riess1998,Perl1999}, cosmic microwave background (CMB) \cite{Spergel:2003cb,Spergel:2006hy,Ade:2013zuv,Ade:2015xua} and baryon acoustic oscillations (BAO) \cite{Eisenstein:2005su,Percival:2009xn},
all imply that the universe is expanding at an increasing rate.
Dark energy (DE) \cite{Copeland:2006wr,Frieman08,Li:2011sd,Li:2013sd} is the most promising way
to explain the accelerating expansion of the universe.
So far, although vast amounts of theoretical DE models have been proposed
\cite{Zlatev:1998tr,ArmendarizPicon:1999rj,Kamenshchik:2001cp,Caldwell02,tpad02,Caldwell:2003vq,Feng:2004ad,hwei05,yxia07,ywang08,Wang:2008te},
the nature of DE is still in dark.

In essence, DE problem may be an issue of quantum gravity. It is commonly believed that the holographic principle (HP)
\cite{Hooft93,suss95} is one of fundamental principles in quantum gravity.
In 2004, Li proposed the so-called holographic dark energy (HDE) model \cite{Li04}, which is the first DE model inspired by the HP.
This model is in very good agreement with current observational data
\cite{Huang:2004zr,Zhang:2005zr,Li:2009zr,Li:2009zs,Huang:2009rf,Wang:2010su,Li:2011wb}
and has drawn a lot of attention in recent 10 years
\cite{Huang:2004ai,Wang:2005jx,Pavon:2005yx,Nojiri:2005pu,LWHZL12,Cui:2015,He:2017,WangLiHu17}.
In addition to HDE, some other HP-inspired DE models have also been proposed,
such as the new agegraphic dark energy (NADE) model \cite{Wei07} and the Ricci dark energy (RDE) model \cite{Gao07}.
For a latest review about the DE models inspired by the HP, see Ref. \cite{WWL17}

As is well known, baryonic matter contains various components.
In addition, dark matter can be mixed by diverse constituents \cite{Bertone:2004pz}.
Therefore, it is reasonable to consider dark energy as a combination of various components. In a recent work \cite{lhde15}, a new DE model called $\Lambda$HDE was proposed, in which DE consists of two parts: cosmological constant $\Lambda$ and HDE.
As far as we know, this is the first theoretical attempt to explore the possibility that DE contains multiple components.
Making use of observational data, Wang et al. constrained parameter space of $\Lambda$CDM, HDE and $\Lambda$HDE \cite{lhde162},
and found that it is difficult to verify whether DE contains one constituent only or not.
Hence, in this paper, we adopt various diagnostic tools to discriminate $\Lambda$HDE from other DE models.

In detail, we diagnose holographic type DE models with three diagnostic tools including the Statefinder hierarchy, composite null diagnostic (CND) and $w-w'$ analysis. The Statefinder hierarchy \cite{Arabsalmani:2011fz} is an upgraded version of the original Statefinder diagnostic \cite{Sahni:2002fz,Alam:2003sc} due to taking higher derivatives of the cosmic scale factor $a(t)$ into account.
CND is a combination of the Statefinder hierarchy and the fractional growth parameter.
$w-w'$ analysis is proposed in \cite{ref_wdw1}, in which the evolution of $w'$ versus the equation of state $w$ is studied.

These three diagnostic tools have been employed to diagnose various DE and modified gravity models \cite{sfh1,sfh2,sfh3}.
However, these research works didn't discuss the $\Lambda$HDE model.
Although Zhou and Wang diagnosed $\Lambda$HDE with the Statefinder hierarchy and CND \cite{Zhou16},
they didn't compare the results of $\Lambda$HDE with other holographic type DE models.
Besides, previous works didn't make use of $w-w'$ analysis, which is also effective on distinguishing different DE models.
Therefore, in this paper, we diagnose the $\Lambda$HDE model and other holographic type DE models
by using not only the Statefinder hierarchy and CND, but also $w-w'$ analysis.
Moreover, we also consider two different cases in diagnostic:
one is adopting different values of $\Omega_{de0}$, and the other is varying an numerical parameter $C$, for the corresponding models.

This paper is organized as follows. In Sec.~\ref{sect:2}and Sec.~\ref{sect:3}, we briefly review a series of holographic type DE models and various diagnostic tools, respectively. In Sec.~\ref{sect:4}, we present the results of diagnosing the $\Lambda$HDE model compared with different holographic type DE models. Conclusions and discussions are given in Sec.~\ref{sect:5}.

\section{Holographic type dark energy models}
\label{sect:2}

For all the diagnostic methods mentioned in this paper, we only focus on the low-redshift region, so we can neglect the weak effects of radiation and curvature terms from the Hubble parameter. Namely, we consider a spatially flat Friedmann-Robertson-Walker (FRW) universe containing matter and DE only. Then, the Friedmann equation takes the form

\begin {equation}
\label{FRW}
H^2=\frac{1 }{3M^2_p}(\rho_m+\rho_{de}),
\end {equation}
where $H=\dot{a}/a$ is the Hubble parameter (the dot denotes the derivative with respect to time $t$), $M_p = (8\pi G)^{-1/2}$ is the reduced Planck mass, $\rho_{m}$ and $\rho_{de}$ are the energy densities for matter and DE, respectively.

According to the holographic principle, the density of HDE is defined as
\begin{equation}
\rho_{de}=3C^2{M_p}^2L^{-2},
\end{equation}
where $C$ is a numerical parameter, and $L$ is the largest infrared (IR) cutoff. Choosing different $L$ will yield different holographic type DE models, and we mainly introduce four types, including the HDE, $\Lambda$HDE, NADE and RDE models, as below.

\subsection{The HDE model}
\label{HDE}

In the original HDE model \cite{Li04}, the density of HDE is $\rho _{de} = 3C^2M^2_pL^{-2}$, and the IR cutoff is taken as the future event horizon given by
\begin{equation}
L=a\int_t^{+\infty}{dt\over a(t)}=a\int_{\it a}^\infty\frac{da'}{Ha'^2}.
\end{equation}
The fractional density of HDE satisfies the differential equation below.
\begin{equation}
\Omega '_{de}= \Omega_{de}(1- \Omega_{de})\left(1+\frac{2}{C}\sqrt{ \Omega_{de}}\right),
\end{equation}
where the prime denotes derivative with respect to $\ln a$.
And the equation of state (EoS) of HDE is given by
\begin{equation}
\label{HDEw}
w\equiv\frac{p_{de}}{\rho_{de}}= -\frac{1}{3}-\frac{2}{3C}\sqrt{ \Omega_{de}}.
\end{equation}

\subsection{The $\Lambda$HDE model}
\label{LHDE}

As mentioned above, the $\Lambda$HDE \cite{lhde15} model consists two constituents.
One is the cosmological constant $\Lambda$, the other is the original HDE part.
Accordingly the energy density of $\Lambda$HDE is given by
\begin{equation}
\rho_{de} =\rho_{\Lambda}+\rho_{hde}=M_{pl}^2\Lambda +3C^2M_p^2L^{-2},
\end{equation}
where the IR cut-off length scale $L$ takes the same form as the HDE model.

The fractional density of $\Lambda$HDE is the solution of the following differential equation:
\begin{equation}
  \frac{d}{dx}\ln{\left|\frac{\Omega_{hde}}{1-\Omega_{hde}}\right|}=\frac{2}{C}\sqrt{\Omega_{hde}}-\frac{d}{dx}\ln{\left|g(a)\right|},
\end{equation}
where $g(a)$ defined by $g(a)\equiv\Omega_{m0}H^{2}_{0}a^{-1}+\Omega_{\Lambda0}H^{2}_{0}a^{2}$, $\Omega_{\Lambda0}={\rho_{\Lambda0}/\rho_{c0}}$ is the initial fractional cosmological constant density, $\Omega_{m0}={\rho_{m0}/ \rho_{c0}}$ is the initial fractional matter density, and $\rho_{c0}=3M_{p}^2{H_0}^2$ is the present critical density of the universe.

According to Dalton's law of partial pressures, the EoS of $\Lambda$HDE is given by
\begin{equation}
w_{de}=\frac{p_{de}}{\rho_{de}}=\frac{p_{hde}+p_{\Lambda}}{\rho_{hde}+\rho_{\Lambda}}=\frac{w_{hde}\Omega_{hde}-\Omega_{\Lambda}}{\Omega_{hde}+\Omega_{\Lambda}}.
\end{equation}
where the EoS of the HDE takes the original form (Eq. \ref{HDEw}), namely,
$w_{hde}=-{1\over3}-{2\over3C}\sqrt{\Omega_{hde}}$.

\subsection{The NADE model}
\label{NADE}

The density of NADE \cite{Wei07} is $\rho _{de} = 3n^2M^2_p\eta ^{-2}$, where $n$ is a numerical constant introduced, and the IR cutoff is chosen as the conformal time given by
\begin{equation}
\eta =\int_0^a\frac{da'}{Ha'^2}.
\end{equation}
The fractional density of NADE is described by the following differential equation:
\begin{equation}
\Omega '_{de} = \Omega_{de}(1- \Omega_{de})\left(3-\frac{2}{na}\sqrt{ \Omega_{de}}\right),
\end{equation}
where the prime denotes derivative with respect to $\ln a$ with the initial condition $\Omega_{de}(z_{ini})=n^2(1+z_{ini})^{-2}/4$ at $z_{ini}=2000$.
And the EoS of NADE is given by
\begin{equation}
w = -1+\frac{2}{3na}\sqrt{ \Omega_{de}}.
\end{equation}

\subsection{The RDE model}
\label{RDE}
The density of RDE \cite{Gao07} is $\rho_{de}=3\alpha M^2_p(\stackrel{\centerdot}{H}+2H^2)$, where $\alpha $ is a dimensionless parameter, and the IR cutoff is related to Ricci scalar curvature given by
\begin{equation}
L= \sqrt{\dfrac{-6}{\mathcal{R}}},
\end{equation}
where the Ricci scalar curvature is $\mathcal{R}=-6(\stackrel{\centerdot}{H}+2H^2)$.
THe fractional density of RDE is
\begin{equation}
\Omega_{de}=\frac{1}{E^2}\left(\frac{\alpha}{2-\alpha}\Omega_{m0}e^{-3x}+f_0e^{-(4-\frac{2}{\alpha})x}\right),
\end{equation}
where $E^2=\Omega_{m0}e^{-3x}+\frac{\alpha}{2-\alpha}\Omega_{m0}e^{-3x}+f_0e^{-(4-\frac{2}{\alpha})x}$ and $f_0=1-\frac{2}{2-\alpha}\Omega_{m0}$ is an integration constant calculated by using the initial condition $E_0=1$.
And the EoS of RDE is given by
\begin{equation}
w=\frac{\frac{\alpha-2}{3\alpha}f_0e^{-(4-\frac{2}{\alpha})x}}{\frac{\alpha}{2-\alpha}\Omega_{m0}e^{-3x}+f_0e^{-(4-\frac{2}{\alpha})x}}.
\end{equation}

\section{The diagnostic methodology}
\label{sect:3}

\subsection{The Statefinder hierarchy}

The Statefinder hierarchy is an impactful geometry diagnostic, which extends and improves the original Statefinder diagnostic by using high-order derivatives of scale factor in order to distinguish diverse DE models more effectively.

We Taylor-expand the the cosmic scale factor around the present epoch $t_0$ aimed at deriving the expression of the Statefinder hierarchy:
\begin{equation}
\frac{a(t)}{a_0}=1+\sum\limits_{\emph{n}=1}^{\infty}\frac{A_{\emph{n}}(t_0)}{n!}[H_0(t-t_0)]^n,
\end{equation}
where
\begin{equation}
A_{\emph{n}}=\frac{a(t)^{(n)}}{a(t)H^n},~~n\in N,
\end{equation}
with $a(t)^{(n)}=d^na(t)/dt^n$.
Note that $A_2 = -q$ is the negative value of the deceleration parameter and $A_3$ represents the original Statefinder parameter $r$ \cite{Chiba:1998tc}

For the $\Lambda$CDM model, we can easily get:
\begin{align}
&A_{2}=1-\frac{3}{2}\Omega_{m},\\
&A_{3}=1,\\
&A_{4}=1-\frac{9}{2}\Omega_{m},\\
&A_{5}=1+3\Omega_{m}+\frac{27}{2}\Omega_{m}^{2},~~\rm{etc.},
\end{align}
In order to let every parameter of the Statefinder hierarchy $S_{\emph{n}}$ remain unity in $\Lambda$CDM model during whole cosmic evolution ($
S_{\emph{n}}|_{\Lambda \rm{CDM}}=1$), we redefine $S_{\emph{n}}$ as
\begin{align}
&S_{2}=A_{2}+\frac{3}{2}\Omega_{m},\\
&S_{3}=A_{3},\\
&S_{4}=A_{4}+\frac{9}{2}\Omega_{m},\\
&S_{5}=A_{5}-3\Omega_{m}-\frac{27}{2}\Omega_{m}^{2},~~\rm{etc.}
\end{align}

Since when the hierarchy number $n$ is greater than or equal to three, $\Omega_{m}=\frac{2}{3}(1+q)$ for $\Lambda$CDM. The expression of Statefinder hierarchy can be rewritten in the following form:
\begin{align}
&S^{(1)}_{3}=A_{3},\\
&S^{(1)}_{4}=A_{4}+3(1+q)\\
&S^{(1)}_{5}=A_{5}-2(4+3q)(1+q),~~\rm{etc.},
\end{align}
where the superscript $(1)$ is to distinguish $S^{(1)}_{\emph{n}}$ from $S_{\emph{n}}$. It is obvious that $S^{(1)}_{\emph{n}}$ is equivalent to one for the $\Lambda$CDM model ($S^{(1)}_{\emph{n}}|_{\Lambda \rm{CDM}}=1$). $S^{(1)}_{3}$ and $S^{(1)}_{4}$ are the main objects to study in this paper. We discriminate and diagnose various holographic type DE models by analyzing the evolution of $S^{(1)}_{3}$ and $S^{(1)}_{4}$ with redshift.

%

\subsection{Composite null diagnostic}
\label{host_properties}

The composite null diagnostic is a combination (CND) of the Statefinder hierarchy $S^{(1)}_{\emph{n}}$ and the fractional growth parameter $\epsilon (z)$ \cite{Acquaviva08} which can also be used as a null diagnostic defined as
\begin{equation}
\label{e}
\epsilon(z)=\frac{f(z)}{f_{\Lambda CDM}(z)},
\end{equation}
where $f(z)=d\ln\delta_{m}/d\ln a$ is the growth rate of the linear density perturbation \cite{Wang:1998gt}. And $\delta_{m}$ is we called the perturbation of the matter density defined as $\delta_m = \delta\rho_m / \rho_m$ which satisfies the following differential equation \cite{Pavlov13}:
\begin{equation}
   \label{deltam}
\ddot\delta_m + 2 {\dot a \over a} \dot\delta_m - {1 \over 2M_p^2}\rho_m\delta_m = 0,
\end{equation}
where the dot denotes the derivative with respect to time $t$. We can write down the one-order and two-order derivatives of $\delta_m$ with respect to time $t$ into the following form:
\begin{equation}
   \label{dotdeltam}
\dot\delta_m = f H \delta_m,
\end{equation}
\begin{equation}
   \label{ddotdeltam}
\ddot\delta_m = (\dot f H + f \dot H + f^2 H^2) \delta_m.
\end{equation}

Substituting Eqs. (\ref{dotdeltam}) and (\ref{ddotdeltam}) into Eq. (\ref{deltam}), we can have the first-order differential equation of $f$:
\begin{equation}
\label{f}
{df \over dz} = \frac{f^2 + 2f - f_1(z)}{1 + z} - \frac{dH / dz}{H}f,
\end{equation}
where $f_1(z) = \rho_m / H^2 = \frac{3\Omega_{m0}(1 + z)^3}{2E^2}$, and the present-day fractional matter density $\Omega_{m0} = 1 - \Omega_{de0}$. Combing with the initial assumed condition $f(z = \infty)=0$, this equation can be numerically solved for different DE models. And after substituting the expression of $f$ back to Eq. (\ref{e}), we get the the fractional growth parameter.
In this paper, we mainly analyze two CND pairs which is \{$S_3^{(1)}, \epsilon$\} and \{$S_4^{(1)}, \epsilon$\}. We plot the trajectories of $S^{(1)}_{\emph{n}}(\epsilon)$ for diverse DE models and in the  $S^{(1)}_{\emph{n}}-\epsilon$ plane, we can directly measure the difference of different DE models.

\subsection{$w-w'$ analysis}
The function $w$ is a state parameter characterizing the dark energy model defined as
$w\equiv{p_{de}/\rho_{de}}$.
and after taking derivative of $w$ with respect to $ln a$, we get
\begin{equation}
w'={dw\over dln a}.
\end{equation}

We analyze the evolution of $w$ and $w'$, and then study the influence of different model parameters on the evolution trajectories of $\{w, w'\}$  \cite{ref_wdw1}. In $w-w'$ plane, the $\Lambda$CDM model is a single point at (-1,0) due to the fact that the state of cosmological constant is equivalent to minus one.

\section{Results}
\label{sect:4}

As mentioned before, we use the Statefinder hierarchy, composite null diagnostic and $w-w' analysis$ to distinguish different holographic type DE models.
In addition, we also compare the evolutionary results of $\Lambda$HDE with $\Lambda$CDM, HDE, NADE and RDE.  In particular, We consider two different cases: (1) varying $\Omega_{de0}$ among 0.5, 0.6, 0.7 and 0.8 while fixing $C=0.6$; (2) varying $C$ among 0.2, 0.4, 0.6 and 0.8 while fixing $\Omega_{de0}=0.7$ (for $\Lambda$HDE, also need fixing $\Omega_{\Lambda 0}=0.4$).

\subsection{Analysis of the Statefinder hierarchy}
\label{asfh}

\begin{figure*}
\includegraphics[width=1.0\textwidth]{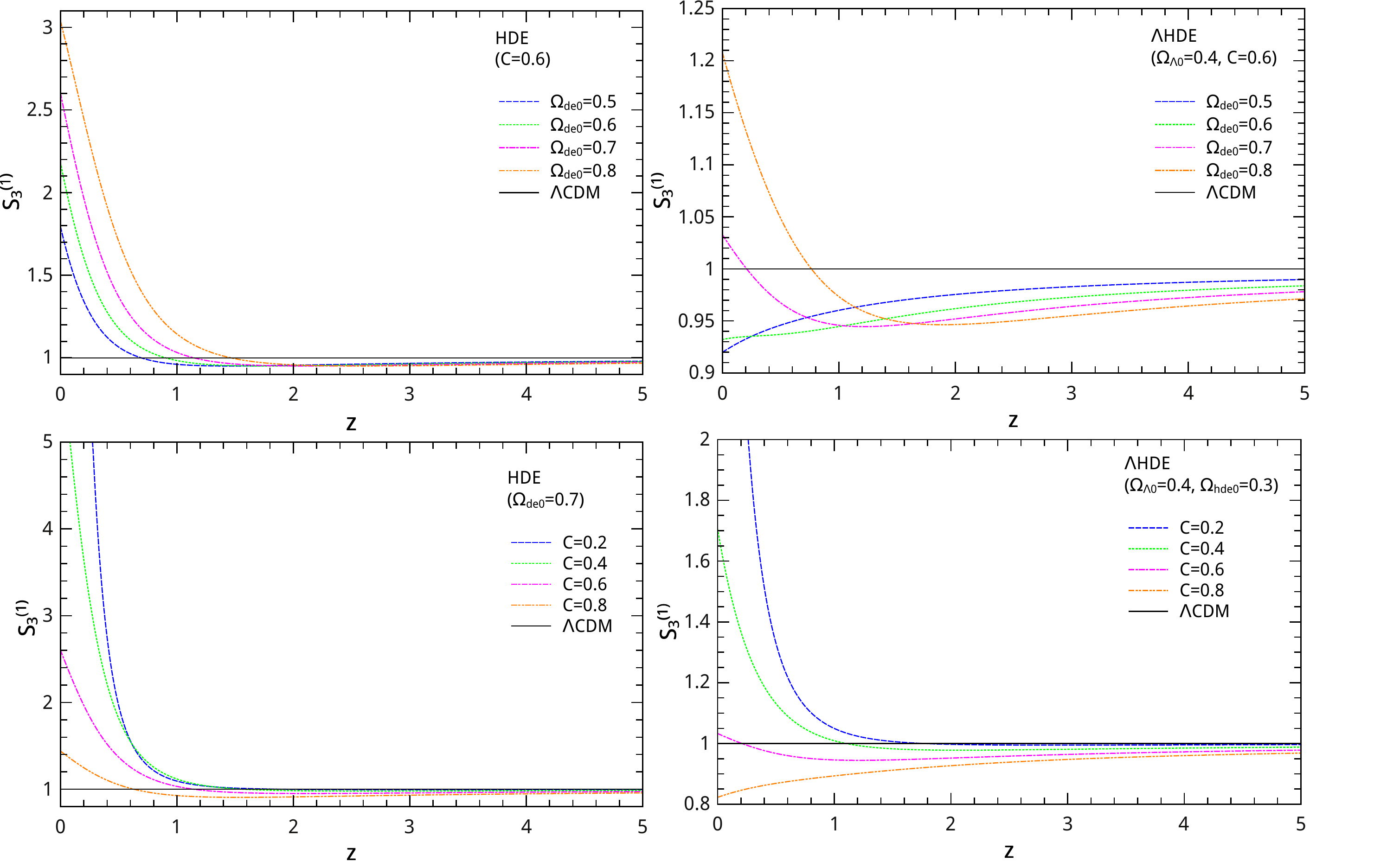}
\caption{\label{S3z} Evolutions of $S^{(1)}_3$ versus redshift $z$ for the HDE model (left panels) and the $\Lambda$HDE model (right panels), with varying $\Omega_{de0}$ (upper panels) and varying $C$ (lower panels). Note that the result of the $\Lambda$CDM is also plotted as a solid horizontal line for comparison.}
\end{figure*}

\begin{figure*}
\includegraphics[width=1.0\textwidth]{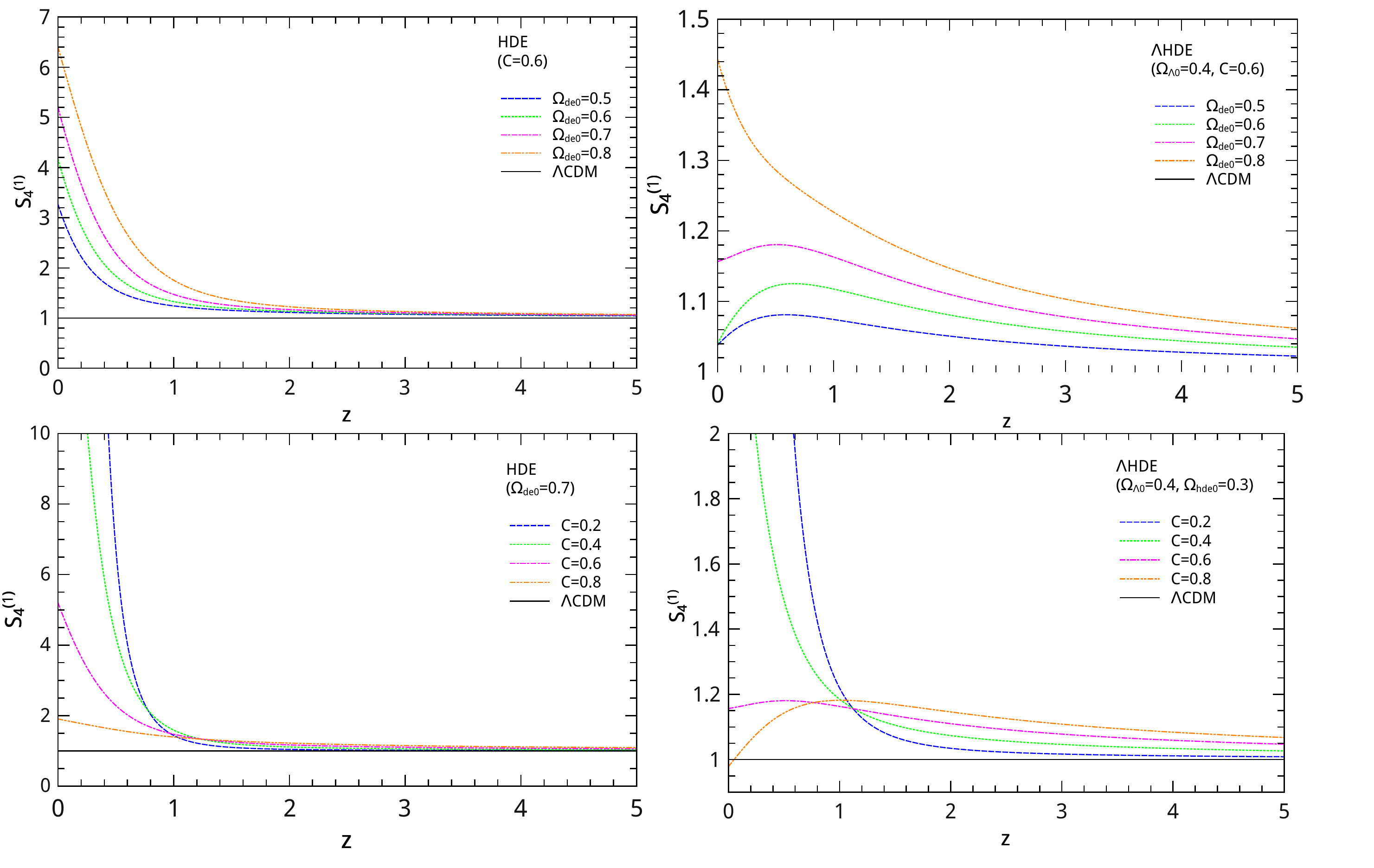}
\caption{\label{S4z} Evolutions of $S^{(1)}_4$ versus redshift $z$ for the HDE model (left panels) and $\Lambda$HDE model (right panels), with varying $\Omega_{de0}$ (upper panels) and varying $C$ (lower panels). Note that the result of the $\Lambda$CDM model is also plotted as a solid horizontal line for comparison.}
\end{figure*}

In this section, we mainly focus on diagnosing $\Lambda$HDE with the Statefinder hierarchy. Moreover, we give the corresponding results of HDE for comparison.

Fig. \ref{S3z} shows the evolutionary trajectories of $S^{(1)}_3(z)$ for HDE (left panels) and $\Lambda$HDE (right panels), with varying $\Omega_{de0}$ (upper panels) and varying $C$ (lower panels).
For the evolution of $S^{(1)}_3(z)$ in the $\Lambda$HDE model, the differentiation of curvilinear shape is more distinct than the original HDE.
Specifically, for the case of varying $\Omega_{de0}$
or $C$ in HDE (left panels), all the curves of $S_3^{(1)}(z)$ have similar evolutionary trajectories, and the trend of curves $S_3^{(1)}(z)$ is monotonic decreasing at the beginning of region of redshift and then closely degenerate into $\Lambda$CDM together. These results indicate that adopting different values of $\Omega_{de0}$ or $C$ only has quantitative impacts on the cosmic evolution of $S_3^{(1)}$ for HDE.
As for $\Lambda$HDE, when
$\Omega_{de0}$ equals to 0.7 and 0.8,
the curves of $S_3^{(1)}(z)$ still have similar evolutionary trend,
but for $\Omega_{de0}=0.5, 0.6$,
the curves of $S_3^{(1)}(z)$ evolves towards an opposite direction (upper-right panel).
In addition, when $C$ increase from 0.6 to 0.8, the evolutions of $S_3^{(1)}(z)$ change into the opposite direction (lower-right panel).
This means that, adopting different values of $\Omega_{de0}$ or $C$ has qualitative impacts on the evolution of $S_3^{(1)}(z)$ for $\Lambda$HDE.

In Fig. \ref{S4z}, we plot the evolutions of $S^{(1)}_4$ versus redshift $z$ for the HDE model (left panels) and the $\Lambda$HDE model (right panels), with varying $\Omega_{de0}$ (upper panels) and varying $C$ (lower panels). The evolutionary trajectories of $S^{(1)}_4(z)$
show similar characteristic as the curves of $S^{(1)}_3(z)$. These results show that adopting different values of $\Omega_{de0}$ has qualitative impacts on the cosmic evolution of the $\Lambda$HDE model. Therefore, as can be seen in Fig. \ref{S3z} and Fig. \ref{S4z}, the differentiation of curvilinear shape between different values of parameters for the $\Lambda$HDE model is more distinct than the original HDE model.

\begin{table*}

\centering
\begin{tabular}{|lcccccccccccccccc|}
\hline
\hline
 &\multicolumn{4}{c}{HDE $(C=0.6)$}&&\multicolumn{4}{c}{$\Lambda$HDE $(\Omega_{\Lambda 0}=0.4, C=0.6)$}&\\
\cline{2-5}\cline{7-10}
$\Omega_{de0}$&$0.5$&$0.6$&$0.7$&$0.8$&&$0.5$&$0.6$&$0.7$&$0.8$&\\
\hline
$S^{(1)}_{3t_{0}}$&$1.79$&$2.18$&$2.60$&$3.04$&&$0.92$&$0.93$&$1.03$&$1.21$&\\
$S^{(1)}_{4t_{0}}$&$3.27$&$4.17$&$5.20$&$6.40$&&$1.04$&$1.04$&$1.16$&$1.44$&\\
\hline
$\Delta S^{(1)}_{3t_{0}}$&\multicolumn{4}{c}{$1.25$}&&\multicolumn{4}{c}{$0.29$}&\\
$\Delta S^{(1)}_{4t_{0}}$&\multicolumn{4}{c}{$3.13$}&&\multicolumn{4}{c}{$0.40$}&\\
\hline
\end{tabular}
\caption{\label{tab1} The present values of the Statefinder hierarchy pairs $S^{(1)}_{3t_{0}}$, $S^{(1)}_{4t_{0}}$ and their maximum differences, $\Delta S^{(1)}_{3t_{0}}$ and $\Delta S^{(1)}_{4t_{0}}$, for HDE and $\Lambda$HDE with varying $\Omega_{de0}$, where $\Delta S^{(1)}_{3t_{0}}=S^{(1)}_{3t_{0}}(\rm{max})-S^{(1)}_{3t_{0}}(\rm{min})$ and $\Delta S^{(1)}_{4t_{0}}=S^{(1)}_{4t_{0}}(\rm{max})-S^{(1)}_{4t_{0}}(\rm{min})$ for each model.}
\end{table*}

\begin{table*}
\centering
\begin{tabular}{|lcccccccccccccccc|}
\hline
\hline
 &\multicolumn{4}{c}{HDE $(\Omega_{de0}=0.7)$}&&\multicolumn{4}{c}{$\Lambda$HDE $(\Omega_{\Lambda 0}=0.4, \Omega_{hde0}=0.3)$}&\\
\cline{2-5}\cline{7-10}
C&$0.2$&$0.4$&$0.6$&$0.8$&&$0.2$&$0.4$&$0.6$&$0.8$&\\
\hline
$S^{(1)}_{3t_{0}}$&$25.99$&$6.10$&$2.60$&$1.44$&&$5.75$&$1.71$&$1.03$&$0.82$&\\
$S^{(1)}_{4t_{0}}$&$278.19$&$24.69$&$5.20$&$1.91$&&$38.55$&$3.28$&$1.16$&$0.98$&\\
\hline
$\Delta S^{(1)}_{3t_{0}}$&\multicolumn{4}{c}{$24.55$}&&\multicolumn{4}{c}{$4.93$}&\\
$\Delta S^{(1)}_{4t_{0}}$&\multicolumn{4}{c}{$276.28$}&&\multicolumn{4}{c}{$37.57$}&\\
\hline
\end{tabular}
\caption{\label{tab2} The present values of the Statefinders hierarchy pairs $S^{(1)}_{3t_{0}}$, $S^{(1)}_{4t_{0}}$ and their maximum differences, $\Delta S^{(1)}_{3t_{0}}$ and $\Delta S^{(1)}_{4t_{0}}$ for HDE and $\Lambda$HDE with varying $C$, where $\Delta S^{(1)}_{3t_{0}}=S^{(1)}_{3t_{0}}(\rm{max})-S^{(1)}_{3t_{0}}(\rm{min})$ and $\Delta S^{(1)}_{4t_{0}}=S^{(1)}_{4t_{0}}(\rm{max})-S^{(1)}_{4t_{0}}(\rm{min})$ for each model.}
\end{table*}

Table \ref{tab1} and  Table \ref{tab2} show the present values of the Statefinder hierarchy pairs, $S^{(1)}_{3t_{0}}$ and $S^{(1)}_{4t_{0}}$, and their maximum differences,$\Delta S^{(1)}_{3t_{0}}$ and $\Delta S^{(1)}_{4t_{0}}$, for HDE and $\Lambda$HDE with varying $\Omega_{de0}$ and varying $C$. Note that $\Delta S^{(1)}_{3t_{0}}=S^{(1)}_{3t_{0}}(\rm{max})-S^{(1)}_{3t_{0}}(\rm{min})$ and $\Delta S^{(1)}_{4t_{0}}=S^{(1)}_{4t_{0}}(\rm{max})-S^{(1)}_{4t_{0}}(\rm{min})$ within each model.
As showing in Table \ref{tab1}, for the case of adopting different values of $\Omega_{de0}$,
we have $\Delta S^{(1)}_{4t_{0}}=3.13>\Delta S^{(1)}_{3t_{0}}=1.25$ for HDE and $\Delta S^{(1)}_{4t_{0}}=0.4>\Delta S^{(1)}_{3t_{0}}=0.29$ for $\Lambda$HDE.
Showing in Table \ref{tab2},
for the case of varying $C$, $\Delta S^{(1)}_{4t_{0}}=276.28>\Delta S^{(1)}_{3t_{0}}=24.55$ for HDE and $\Delta S^{(1)}_{4t_{0}}=37.57>\Delta S^{(1)}_{3t_{0}}=4.93$ for $\Lambda$HDE. The data from these tables \ref{tab1} and \ref{tab2} all indicates that $\Delta S_{4t_{0}}^{(1)}$ is remarkably larger than $\Delta S_{3t_{0}}^{(1)}$. Therefore, compared with $S_3^{(1)}$, $S_4^{(1)}$ can give larger differences among the cosmic evolutions of the holographic type DE models associated with different $\Omega_{\Lambda0}$ or different $C$ which makes us easier to distinguish different theoretical models.

\subsection{Analysis of the composite null diagnostic}
\label{acnd}

\begin{figure*}
\includegraphics[width=1.0\textwidth]{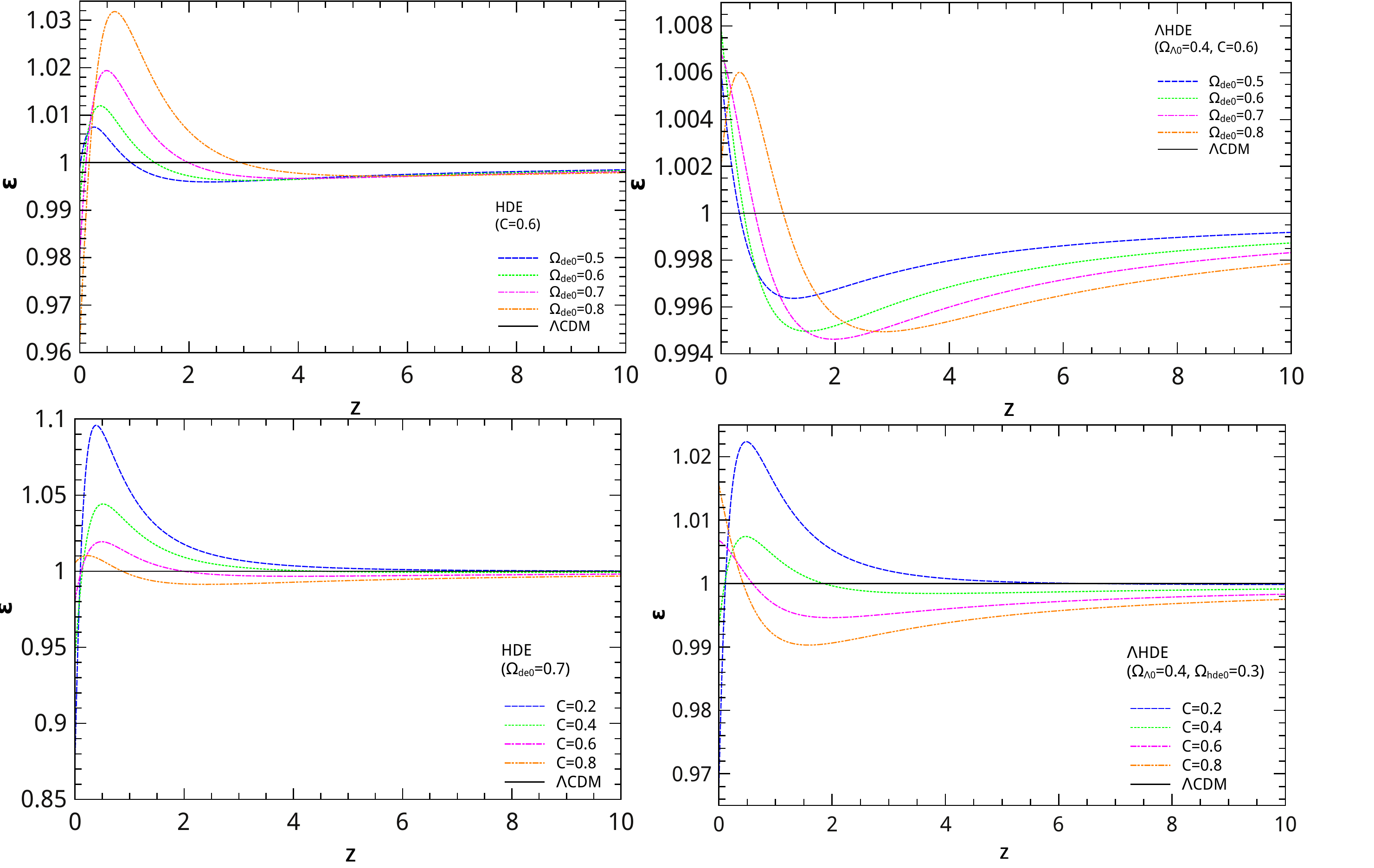}
\caption{\label{ez} Evolutions of $\epsilon$ versus redshift $z$ for the HDE model (left panels) and $\Lambda$HDE model (right panels), with varying $\Omega_{de0}$ (upper panels) and varying $C$ (lower panels). Note that the result of the $\Lambda$CDM model is also plotted as a solid horizontal line for comparison.}
\end{figure*}

\begin{figure*}
\includegraphics[width=1.0\textwidth]{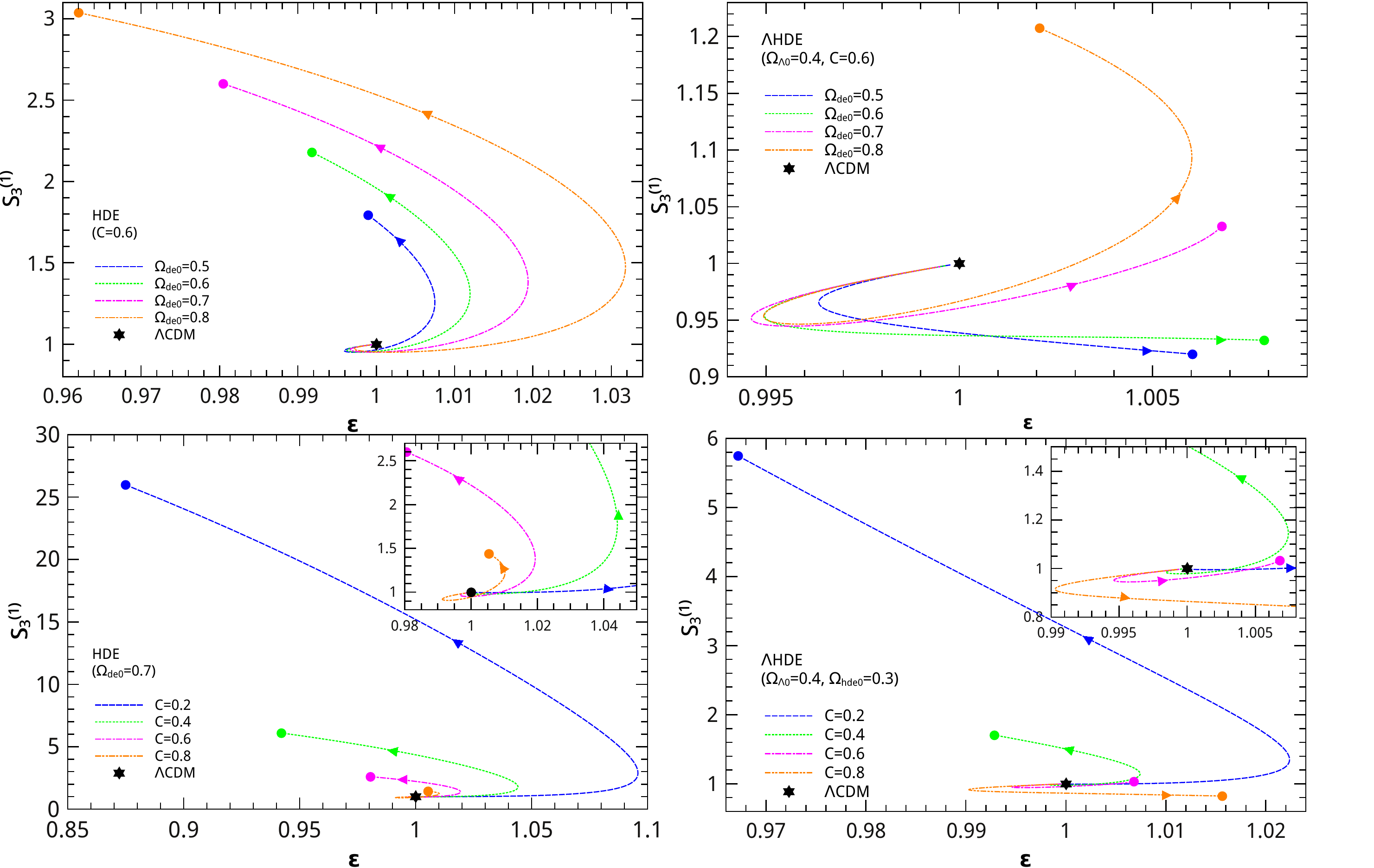}
\caption{\label{S3e} The evolutionary trajectories of the CND pair \{$S_3^{(1)}, \epsilon$\} for the HDE model (left panels) and the $\Lambda$HDE model (right panels), with varying $\Omega_{de0}$ (upper panels) and varying $C$ (lower panels). The current values of \{$S_3^{(1)}, \epsilon$\} for the HDE and $\Lambda$HDE are marked by solid dots. For the $\Lambda$CDM model, $\{S^{(1)}_3, \epsilon\}=\{1,1\}$ is marked by hexagram for comparison. Note that the arrows indicate the time directions of cosmic evolution ($z \to 0$).}
\end{figure*}

\begin{figure*}
\includegraphics[width=1.0\textwidth]{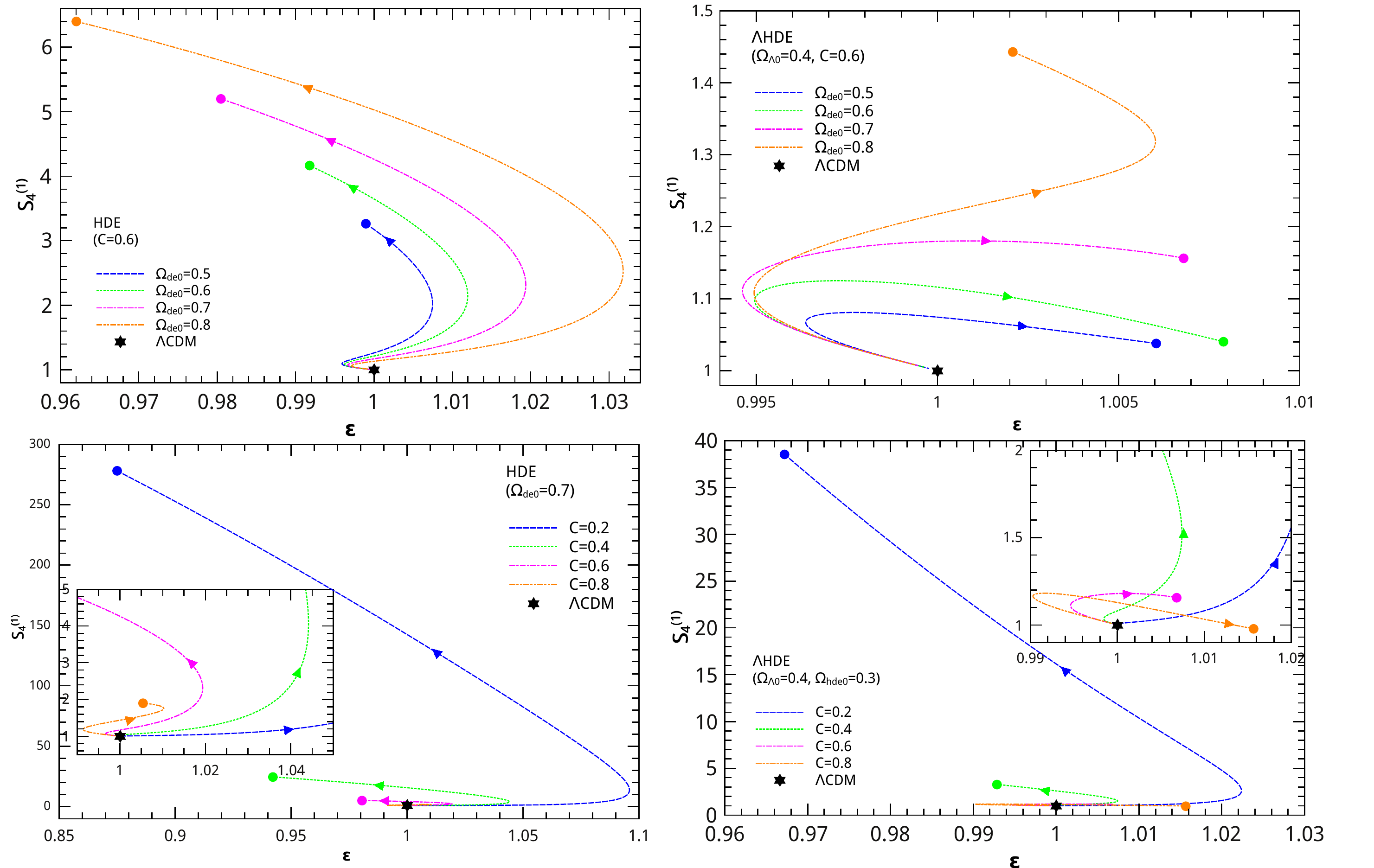}
\caption{\label{S4e} The evolutionary trajectories of the CND pair \{$S_4^{(1)}, \epsilon$\} for the HDE (left panels) and $\Lambda$HDE (right panels), with varying $\Omega_{de0}$ (upper panels) and varying $C$ (lower panels). The current values of \{$S_4^{(1)}, \epsilon$\} for the HDE model and the $\Lambda$HDE model are marked by solid dots. For the $\Lambda$CDM model, $\{S^{(1)}_4, \epsilon\}=\{1,1\}$ is marked by hexagram for comparison. Note that the arrows indicate the time directions of cosmic evolution ($z \to 0$).}
\end{figure*}

Since using one single diagnostic tool can analyze and present one-side information of cosmic evolution only, applying CND can make use of both geometrical and matter perturbational information of cosmic evolution.
To diagnose diverse theoretical DE models with the CND pairs, \{$S_3^{(1)}, \epsilon$\} and \{$S_4^{(1)}, \epsilon$\}, first we analyse the evolution of the fractional growth parameter $\epsilon(z)$.

Fig. \ref{ez} is the evolutionary trajectories of $\epsilon$ versus redshift $z$ for HDE (left panels) and $\Lambda$HDE (right panels), with varying $\Omega_{de0}$ (upper panels) and varying $C$ (lower panels).
The evolutionary trajectories of $\epsilon(z)$ has similar characteristic as the curves of $S^{(1)}_3(z)$ and $S^{(1)}_4(z)$. For instance, while adopting different $\Omega_{de0}$, all the curves $\epsilon(z)$ of HDE have convex vertices at low-redshift region and descend monotonically at higher-redshift (left panels), but for $\Lambda$HDE, the curves of $\epsilon(z)$ change from the shape of having convex and concave vertices to having concave vertices only while decreasing the values of $\Omega_{de0}$ (right panels). These results show that adopting different values of $\Omega_{de0}$ or $C$ has qualitative impacts on the cosmic evolution of the $\Lambda$HDE model. As can be seen in Fig. \ref{S3z}, Fig. \ref{S4z} and Fig. \ref{ez}, for $\Lambda$HDE, the differentiation of curvilinear shape between different values of parameters is more distinct than the original HDE model.

Fig. \ref{S3e} and Fig. \ref{S4e} are using the CND method to distinguish different DE models. Except analysing the Statefinder hierarchy pairs $S_3^{(1)}$ and $S_4^{(1)}$, CND method make use of fractional growth parameter $\epsilon$ at the same time. As we have shown before, the trajectories of $\epsilon(z)$ have different evolution while adopting different values of $\Omega_{de0}$ or $C$.
This means that, during the cosmic evolution, CND method contains more information of difference between varying parameter. Compared with $S_3^{(1)}(z)$ (Fig. \ref{S3z}), $S_4^{(1)}(z)$ (Fig. \ref{S4z}) and $\epsilon(z) (Fig. \ref{ez})$, CND pairs have significantly different evolutionary trajectories (Fig. \ref{S3e} and Fig. \ref{S4e}), which can be used to distinguish various DE models more evidently.

In Fig. \ref{S3e}, we plot the evolution of the CND pair \{$S_3^{(1)}, \epsilon$\} for HDE (left panels) and $\Lambda$HDE (right panels), with varying $\Omega_{\Lambda0}$ (upper panels) and varying $C$ (lower panels). Specifically, for the cases of varying $\Omega_{de0}$ or $C$ in HDE (right panels), the curves of CND pairs only have quantitative differences:
at high-redshift region, all curves of \{$S_3^{(1)}, \epsilon$\}
are close to the shape of reverse spiral which starts from the neighbourhood of the hexagram symbol for $\Lambda$CDM, then evolves towards the direction of the decrease of $\epsilon$ and $S_3^{(1)}$, after passing a turning point, it continues evolving towards the direction of the increase of $\epsilon$ and the increase of $S_3^{(1)}$ (left panels).
On the other side, for $\Lambda$HDE, the curves of CND pairs \{$S_3^{(1)}, \epsilon$\} have qualitative differences:
the curve for $\Omega_{de0}=0.8$ (upper-right panel) has an analogous shape of evolutionary trajectory in HDE. In contrast, while $\Omega_{de0}$ decreasing to 0.5 and 0.6, the trajectories evolve towards to the direction of increase $eplison$ and decrease of $S_3^{(1)}$ after passing the turning point;
For the case of varying $C$ in $\Lambda$HDE (lower-right panel), the curves \{$S_3^{(1)}, \epsilon$\} also have the similar characteristic as the case of varying $\Omega_{de0}$.
These results further verify that for $\Lambda$HDE, the curves of CND pairs \{$S_3^{(1)}, \epsilon$\} have qualitative differences while varying the values of $\Omega_{de0}$ or $C$.

Fig. \ref{S4e} is the the evolutionary trajectories of the CND pair \{$S_4^{(1)}, \epsilon$\} for HDE (left panels) and $\Lambda$HDE (right panels), with varying $\Omega_{\Lambda0}$ (upper panels) and varying $C$ (lower panels). Similarly, the same characteristic is still workable for the evolutionary trajectories of CND pairs \{$S_4^{(1)}, \epsilon$\}: for the $\Lambda$HDE model, the curves of CND pairs \{$S_4^{(1)}, \epsilon$\} with varying $\Omega_{de0}$ or $C$ have qualitative differences, which is more distinct than the differentiation of curvilinear shape in HDE.

\begin{figure*}
\includegraphics[width=1.0\textwidth]{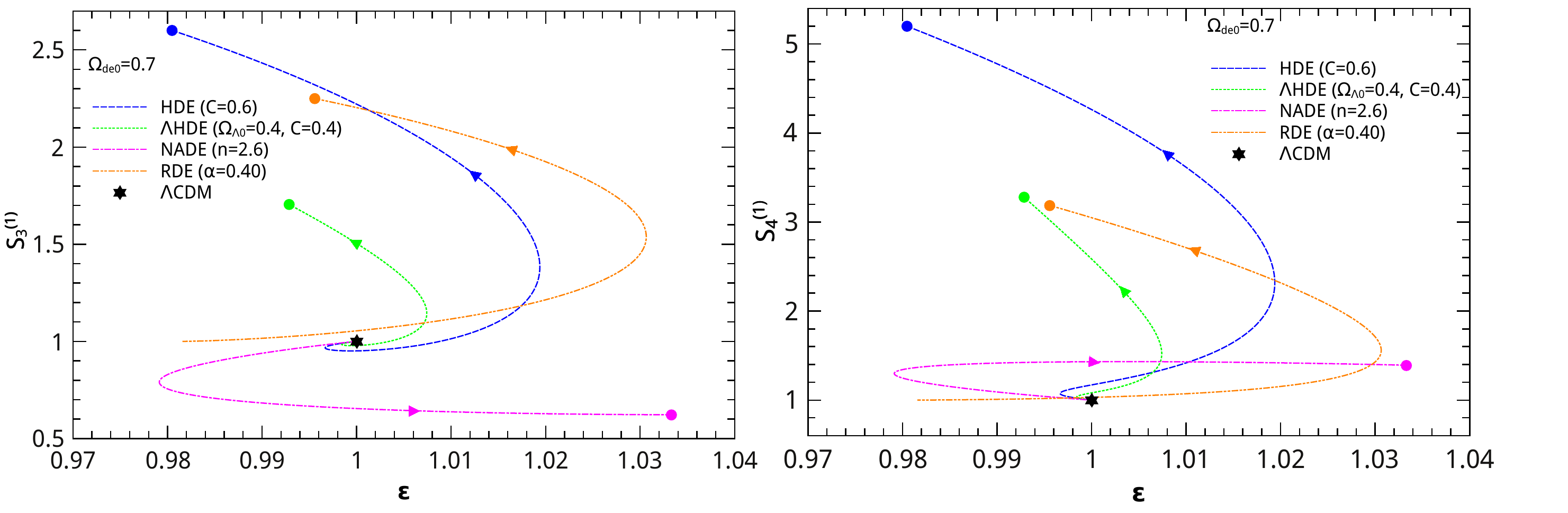}
\caption{\label{fig6} Comparisons of the evolutionary trajectories of $\{S^{(1)}_3, \epsilon\}$ and $\{S^{(1)}_4, \epsilon\}$ for the HDE, $\Lambda$HDE, NADE and RDE models in the $S^{(1)}_3$--$\epsilon$ plane.
The current values of $\{S^{(1)}_3, \epsilon\}$ and $\{S^{(1)}_4, \epsilon\}$ for the holographic type DE models are marked by the round dots.
$\{S^{(1)}_3, \epsilon\}=\{1,1\}$ and $\{S^{(1)}_4, \epsilon\}=\{1,1\}$ for the $\Lambda$CDM model is also shown as hexagram for comparison. The arrows indicate the evolution directions of the models.}
\end{figure*}

For particluar analysis of Statefinder hierarchy for NADE and RDE, see \cite{sfh3}. To show the relative trend and location of evolution trajectories for $\Lambda$HDE compared with other holographic type DE models, we add NADE and RDE for comparison. As shown in Fig. \ref{fig6}, we compare the evolution trajectories of $\{S^{(1)}_3, \epsilon\}$ and $\{S^{(1)}_4, \epsilon\}$ for HDE, $\Lambda$HDE, NADE and RDE. For all models except single-parameter NADE model, we fix $\Omega_{de0}=0.7$ and choose typical values of the parameters which are close to the current observational constraints of each model. In HDE, the parameter $C$ takes 0.6 \cite{Wang:2012uf}. In $\Lambda$HDE, $\Omega_{\Lambda0}$ is 0.4 and $C$ takes 0.6 \cite{lhde15}. For NADE which is a single-parameter model with the initial condition $\Omega_{de}(z_{ini})=n^2(1+z_{ini})^{-2}/4$ at $z_{ini}=2000$ \cite{Wei:2007xu}, we choose $n=2.6$ \cite{Zhang:2013lea}. In RDE, $\alpha$ takes 0.40 \cite{Zhang:2009un}. For $\Lambda$HDE, we can also vary the value of $\Omega_{\Lambda0}$ and meanwhile fix the values of $\Omega_{de0}$ and $C$. For detailed analysis of Statefinder hierarchy for this case, see previous work of our group, and $w'-w$ analysis for this case see Sec. \ref{aww}.

\subsection{Analysis of the $w'-w$ diagnostic}
\label{aww}

\begin{figure*}
\includegraphics[width=1.0\textwidth]{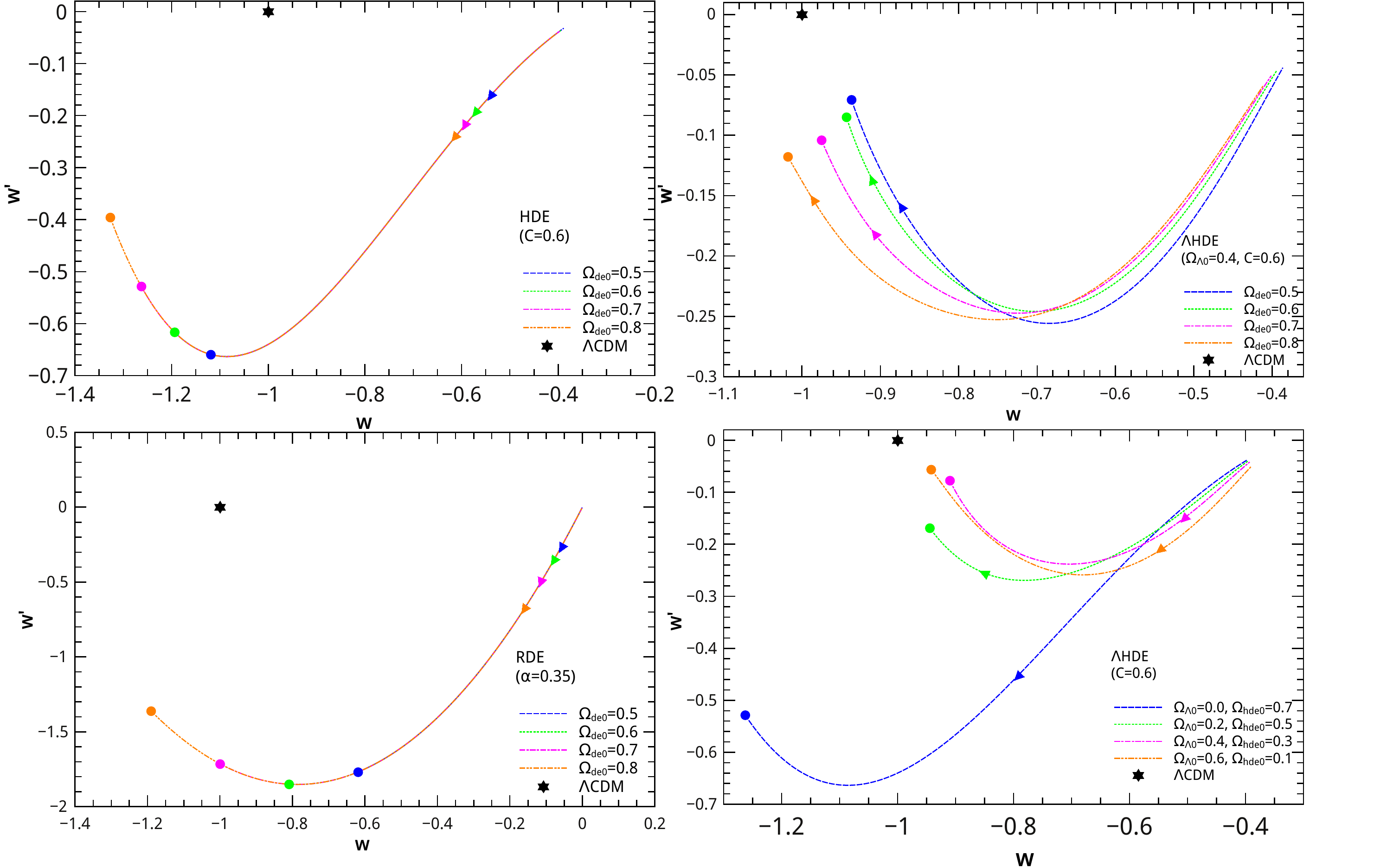}
\caption{\label{ww'} The evolutionary trajectories of the $w'-w$ analysis for holographic type DE models, with varying $\Omega_{\Lambda0}$ and other parameters fixed. The current values of $\{w, w'\}$ are marked by the round dots. For the $\Lambda$CDM model, $\{w, w'\}=\{-1,0\}$ is marked by hexagram for comparison. Note that the arrows indicate the time directions of cosmic evolution ($z \to 0$).}
\end{figure*}

\begin{figure*}
\includegraphics[width=1.0\textwidth]{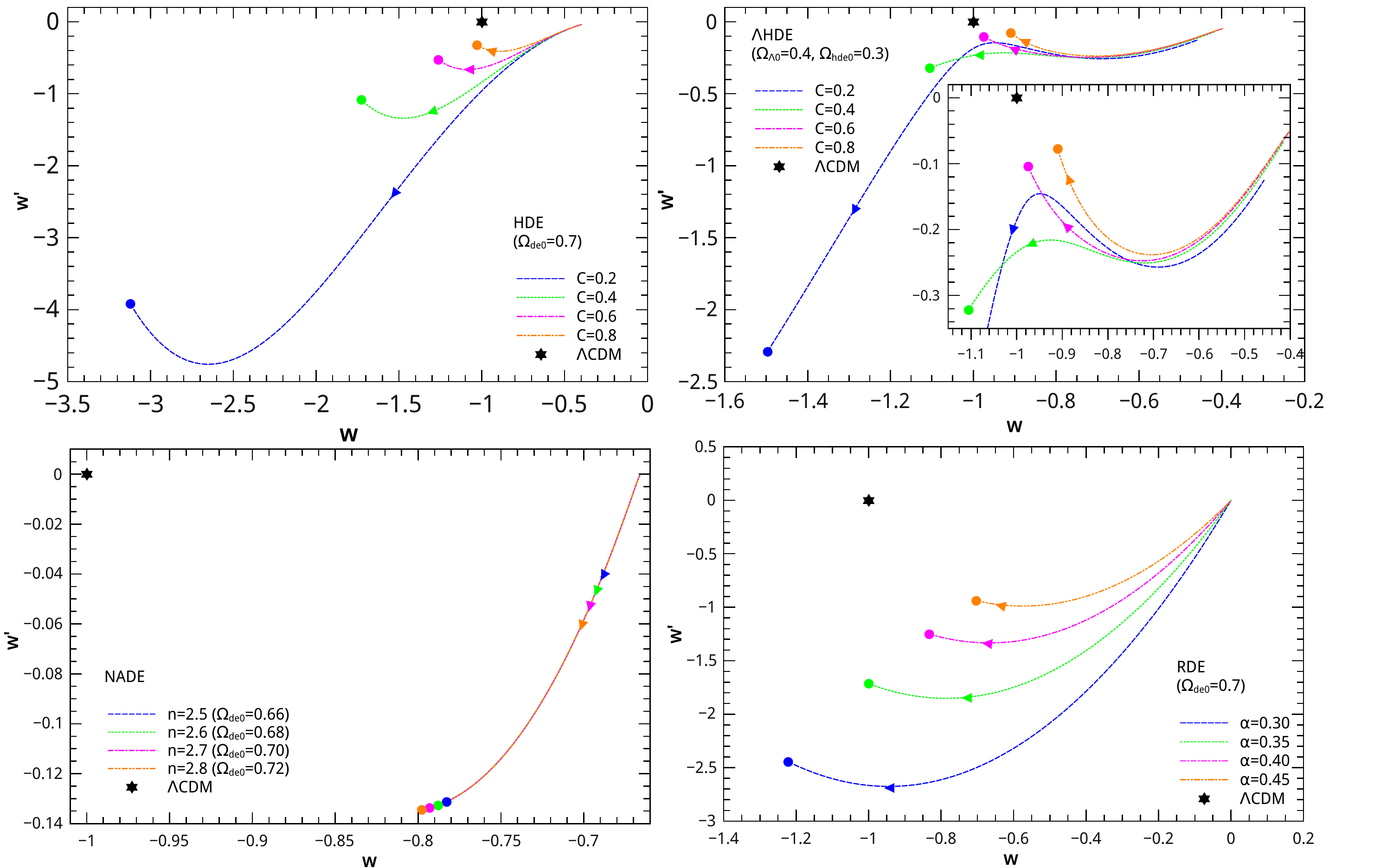}
\caption{\label{ww'c} The evolutionary trajectories of the $w'-w$ analysis for holographic type DE models, with varying numerical parameter ($C$ for HDE and $\Lambda$HDE; $n$ for NADE; $\alpha$ for RDE) and other parameters fixed. The current values of $\{w, w'\}$ are marked by the round dots. For the $\Lambda$CDM model, $\{w, w'\}=\{-1,0\}$ is marked by hexagram for comparison. Note that the arrows indicate the time directions of cosmic evolution ($z \to 0$).}
\end{figure*}

We show in Fig. \ref{ww'} the evolutionary trajectories of the $w'-w$ analysis for holographic type DE models, with varying $\Omega_{de0}$ and other parameters fixed. Since NADE is a single-parameter model by applying the initial condition $\Omega_{de}(z_{ini})=n^2(1+z_{ini})^{-2}/4$ at $z_{ini}=2000$ \cite{Wei:2007xu}, when parameter $n$ takes different particular values, $\Omega_{de0}$ get different values respectively. Unlike the other two-parameter holographic type model, NADE can't fix parameter $\Omega_{de0}$ and meanwhile vary parameter $n$ or vice versa. Note that $n$ takes 2.5, 2.6, 2.7 and 2.8 approximately correspond to the $\Omega_{de0}$ 0.66, 0.68, 0.70 and 0.72 respectively. Therefore we only plot the case of varying parameter $n$ for NADE in lower-left panel of Fig. \ref{ww'c} which can also represent the result of varying parameter $\Omega_{de0}$. In addition, we also plot the evolutionary trajectories of the $w'-w$ analysis for the case of varying the value of $\Omega_{\Lambda0}$ among 0, 0.2, 0.4 and 0.6 and meanwhile fixing the values of $\Omega_{de0}=0.7$ and $C=0.6$ in FIg. \ref{ww'} (lower-right panel).

For HDE, NADE and RDE, the trajectories of $\{w, w'\}$ overlap together substantially during the whole evolution history while varying $\Omega_{de0}$ 
In contrast, as shown in upper-right panel of Fig. \ref{ww'}, adopting different values of $\Omega_{de0}$ has quantitative impacts on the cosmic evolution of the $\Lambda$HDE model which is significantly different with other holographic type DE models. Specifically, the trajectories of $\{w, w'\}$  have a trend of coincidence in high-redshift region but this trend of coincidence is separated with the decrease of redshift.
As for the case of varying  $\Omega_{\Lambda0}$, the curves of $\{w, w'\}$ has quantitative impacts during the cosmic evolution for $\Lambda$HDE (lower-right panels). Note that the $\{w, w'\}$ curves of $\Lambda$HDE associated with $\Omega_{\Lambda0}\neq0$ is clearly distinguished from the particular curve with $\Omega_{\Lambda0}=0$ in which $\Lambda$HDE degenerate into HDE.
Therefore, these results show the uniqueness of $\Lambda$HDE in $w'-w$ analysis compared with other mono-component holographic type DE models.

As shown in Fig .\ref{ww'c}, we plot the evolution of $w'$ versus $w$ for the HDE (left panels) and $\Lambda$HDE (right panels), with varying numerical parameter ($C$ for HDE and $\Lambda$HDE; $n$ for NADE; $\alpha$ for RDE) and other parameters fixed. Since NADE is a single-parameter model, the evolutionary trajectories of NADE are overlap together substantially which is similar to the case of adopting different  $\Omega_{de0}$ for HDE and RDE in Fig .\ref{ww'}. For both HDE and RDE, the curves have a trend of coincidence in high-redshift region, but have quantitative differences in low-redshift region (lower panels). For $\Lambda$HDE (upper-right panel), the evolution also degenerate in high-redshift but showing qualitative differences between different values of $C$ in low-redshift, which can be a specific property to distinguish other mono-component DE models. In detail, the curves associated with $C=0.2$ and $C=0.4$ are in different evolutionary shape compared with the curves of $C=0.6$ and $C=0.8$ in low-redshift region.

\section{Conclusion and Discussion}
\label{sect:5}
In this paper, we diagnose holographic type DE models with the Statefinder hierarchy, composite null diagnostic and $w-w'$ analysis. In particular, we diagnose $\Lambda$HDE which is a bi-component DE model compared with other mono-component DE models.

Specifically, for the case of varying $\Omega_{de0}$ or $C$ in HDE, all the trajectories of $S_3^{(1)}(z)$, $S_4^{(1)}(z)$, $\epsilon(z)$, $\{S_3^{(1)}, \epsilon$\} and $\{S_4^{(1)}, \epsilon$\} have same shape and similar evolutionary trend.
In contrast, adopting different values of $\Omega_{de0}$ or $C$ has qualitative impacts for $\Lambda$HDE.
This means that, the differentiation of curvilinear shape between different values of parameters for $\Lambda$HDE
is more distinct than the original HDE model.

In addition, $S_4^{(1)}$ can always give larger differences among the cosmic evolutions of the holographic type DE models than $S_3^{(1)}$, which makes us easier to distinguish different theoretical models. Moreover, compared with $S_3^{(1)}(z)$, $S_4^{(1)}(z)$ and $\epsilon(z)$, CND pair has significantly different evolutionary trajectories, which are more effective on diagnosing diverse theoretical DE models.

Moreover, the results of $w'-w$ analysis for holographic type DE models show the special feature of $\Lambda$HDE. To be specific, for HDE, NADE and RDE,the trajectories of $\{w, w'\}$ overlap together substantially during the whole evolution history while varying $\Omega_{de0}$. In contrast, adopting different values of $\Omega_{de0}$ or $\Omega_{\Lambda0}$ has quantitative impacts on the cosmic evolution of $\Lambda$HDE, which is significantly different with other mono-component holographic type DE models.
For the case of varying numerical parameter, the evolutionary trajectories of NADE are overlap together.
For both HDE and RDE, the curves have quantitative differences in low-redshift region.
As for $\Lambda$HDE, the evolution shows qualitative differences in low-redshift,
which can be viewed as a specific property to distinguish other mono-component DE models.

Therefore, we can conclude that compared with HDE, RDE and NADE, the $\Lambda$HDE model can be easily distinguished by using these diagnostic tools.

Recently, it is found that the redshift-evolution of supernova color-luminosity parameter $\beta$ \cite{WangWang2013,Wang:2013mha}
may change the fitting-results of parameter estimation for various cosmological models \cite{Wang:2013tic,WWGZ14,WWZ14,WGHZ15};
moreover, this systematic error of supernova may have significant effect on the studies of various cosmological problems \cite{WHLL16,LLWZ16,HLLW16,WangWen17,WenWang17}.
Therefore, it would be interesting to revisit the $\Lambda$HDE model by taking into account this new factor.
This will be studied in future works.

\section*{Acknowledgements}
We thank Lanjun Zhou and Nan Li for valuable suggestions and technological supports.
SW is supported by the National Natural Science Foundation of China under Grant No. 11405024
and the Fundamental Research Funds for the Central Universities under Grant No. 16lgpy50.

\bibliographystyle{plainnat}

\end{multicols}

\end{document}